\documentclass[
aps,
pra,
twocolumn,
groupedaddress,
showpacs
]{revtex4-1}

\usepackage{graphicx}
\usepackage{dcolumn}
\usepackage{bm}
\usepackage{amsmath}

\begin{document}


\title{Site-resolved imaging of ytterbium atoms in a two-dimensional optical lattice}

\author{Martin Miranda}
\email[]{miranda.m.aa@m.titech.ac.jp}

\author{Ryotaro Inoue}
\author{Yuki Okuyama}
\author{Akimasa Nakamoto}
\author{Mikio Kozuma}

\affiliation{Department of Physics, Tokyo Institute of Technology, 2-12-1 O-okayama, Meguro-ku, Tokyo 152-8550, Japan}

\date{\today}

\begin{abstract}
We report a high-resolution microscope system for imaging ultracold ytterbium atoms trapped in a two-dimensional optical lattice. By using the ultraviolet strong transition combined with a solid immersion lens and high-resolution optics, our system resolved individual sites in an optical lattice with a 544-$\mathrm{nm}$ spacing. Without any cooling mechanism during the imaging process, the deep potential required to contain the atoms was realized using a combination of a shallow ground-state and a deep excited-state potentials. The lifetime and limitations of this setup were studied in detail.

\end{abstract}

\pacs{37.10.Jk, 37.10.Gh, 67.85.Hj, 07.60.Pb}


\maketitle

\section{Introduction}

Quantum gas microscopes are high-resolution fluorescence imaging devices capable of resolving individual atoms trapped in a two-dimensional optical lattice \cite{nature09378, nature08482}. The highly controllable optical lattice potential shape combined with a high-resolution microscope system provides a fascinating tool for simulating quantum systems, such as observing superfluid-to-Mott-insulator phase transitions \cite{Bakr30072010}, quantum magnetism in the Bose-Hubbard regime \cite{simon2011quantum}, quantum dynamics of superexchanging interactions \cite{chen2011controlling}, and strongly correlated quantum phenomena \cite{Preiss13032015}. Currently, researchers are working to develop quantum gas microscopes for several atomic species \cite{PhysRevLett.114.213002, PhysRevLett.114.193001, haller2015single}, which would allow them to simulate a wider diversity of systems, such as strongly correlated Fermi-Hubbard systems. 

In this paper, we report a system to address ultracold ytterbium atoms trapped in a two-dimensional optical lattice with single-site resolution. The Yb atom, having no electronic spin in the ground state, is robust against decoherence due to magnetic-field fluctuations and therefore is a promising atomic species for quantum information processing \cite{PhysRevLett.101.170504, PhysRevLett.102.110503}. Another advantage of Yb is that it has a variety of stable isotopes: five bosonic and two fermionic isotopes. One of those isotopes (${}^{171}$Yb) has a 1/2 nuclear spin and is ideal for implementing a quantum bit \cite{takei2010faraday, springerlink:10.1007/s00340-009-3696-4, PhysRevA.84.030301, PhysRevA.81.062308}. Polarization-dependent fluorescence detection combined with the quantum gas microscope would enables us to directly read the spin state of the system \cite{PhysRevLett.106.160501}. Additionally,  Yb has two long-lived metastable states, one of which is useful to tune inter-atomic interactions \cite{PhysRevLett.110.173201}. The wide variety of isotopes also makes possible the study of rich Mott insulator systems \cite{sugawa2011interaction}.

The conventional method to observe atoms trapped in an optical lattice is to excite the atoms and observe the resulting fluorescence using a high-numerical-aperture microscope. Due to the large number of photons required to obtain a well-resolved image, a deep lattice potential is necessary to keep the heated atoms trapped during the imaging process. In the case of rubidium atoms, the polarization- gradient-cooling technique \cite{nature08482, nature09378, PhysRevA.59.R19} worked effectively to overcome this difficulty, as it could cool down the atoms to sub-Doppler temperatures while the resultant fluorescence was observed. Unfortunately, this technique is not effective for all the atomic species. In the case of bosonic Yb atoms, for example, the lack of hyperfine splitting makes impossible the application of sub-Doppler cooling mechanisms such as polarization-gradient-cooling, Raman cooling \cite{PhysRevLett.80.4149, PhysRevLett.81.5768, PhysRevA.72.043409, haller2015single, PhysRevLett.114.213002}, and electromagnetically induced transparency cooling \cite{morigi2000ground, morigi2003cooling, PhysRevLett.114.193001}.

One possible idea to cool Yb atoms would be to use the conventional Doppler cooling at either of the two cooling transitions: the strong ${}^1S_0$-${}^1P_1$ transition with a Doppler cooling limit of 690$\,\mathrm{\mu K}$ and the ${}^1S_0$-${}^3$P${}_1$ intercombination transition having a Doppler cooling limit of 4.4$\,\mathrm{\mu K}$. The latter could, in principle, allow us to trap the atoms with potential depths of approximately $100$$\,\mathrm{\mu K}$. However, the small laser detuning required to achieve low temperatures is sensitive to spatial inhomogeneities in the light shift. As a consequence, such a narrow transition would require the suppression of the inhomogeneities of the optical transition frequency in the optical trap, for example, by using a magic wavelength \cite{katori1999optimal}. 

\begin{figure}[t]
\includegraphics{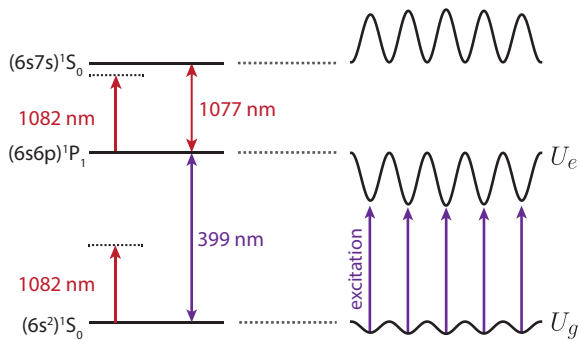} 
\caption{\label{fig:potential} (Color online) The lattice light is near resonant with the upper ${}^1P_1$-${}^1S_0$ transition, creating a large light shift in the  ${}^1P_1$ state. The excitation beam is resonant to the shifted ${}^1S_0$-${}^1P_1$ transition. }
\end{figure}

In our experiment, instead of applying laser cooling, we opted to create a deep potential capable of containing the heated atoms and use the strong ${}^1S_0$-${}^1P_1$ transition (wavelength 399$\,\mathrm{nm}$ and natural linewidth 29$\,\mathrm{MHz}$) to excite the atoms. Taking into account the $10^4$ photons per atom obtained in previous experiments \cite{nature09378, nature08482}, the heating due to this number of emissions corresponds to a temperature of 3.5$\,\mathrm{mK}$, which means that a potential depth of several tens of $\mathrm{mK}$ is required to stably trap the atoms during imaging process. To create such a deep potential, the optical lattice wavelength was selected such that the potential in the excited state $U_e$ is deep and the one in the ground state $U_g$ is shallow. The excitation light couples these states to provide a deep effective potential, as shown in Fig.~\ref{fig:potential}. For our experiment we choose a wavelength of $\lambda_{\text{lat}} = 1082 \,\mathrm{nm}$. This wavelength is red detuned by only $\Delta_{lat}/2\pi=-1.2$$\,\mathrm{THz}$ to the upper ${}^1P_1$-${}^1S_0$ transition (wavelength $1077.2\,\mathrm{nm}$, natural linewidth 3$\,\mathrm{MHz}$), which creates the desired potential shape. More specifically, for this wavelength, the light shift in the excited ${}^1P_1$ state is approximately 200 times larger than that in the ground state. The excitation light is resonant to the light-shifted strong transition ${}^1S_0$-${}^1P_1$.

To see the advantages of this system, we consider first the case in which the excitation light is not present in the system. As the lattice light is far off resonant from any of the transitions from the ground state, the shallow potential $U_g$ results in small scattering rates ($10^{-3}\,\mathrm{s}^{-1}$ for $U_g/k_B = -2\,\mathrm{\mu K}$). During the imaging process, the resonant excitation light illuminates the atoms and couples the lower ${}^1S_0$ and ${}^1P_1$ states. Atoms then experience an average potential $(U_g+U_e)/2$ between the deep excited-state potential and the shallow ground-state one \cite{dalibard1985dressed}. This deep potential is useful to trap the atoms during the imaging process. For this approximation to be true, the Rabi frequency should be much larger than the trap frequency in the optical lattice. The strong transition with saturated intensity satisfies this condition. Note that creating a deep potential not in the excited state but in the ground one would not produce the same desired conditions. As the only strong transition from the ground state is the ${}^1S_0$-${}^1P_1$ transition, creating a deep potential would require using a lattice light which is near resonant to this transition. In this case, even for small intensities the scattering rates in the ground state would be on the order of $10\,\mathrm{s}^{-1}$ for potential depths of a few $\mathrm{\mu K}$, which is not a good condition for creating a Bose-Einstein condensate (BEC) and a Mott insulator. Also, as the wavelengths of the lattice light and the excitation light are similar, it is difficult to reduce the effects of stray lights. Additionally, high-power laser sources at this wavelength are not available.

This paper is organized as follows. 
Section \ref{section:experiment} briefly presents the experimental setup including the method to load ultracold Yb atoms into a two-dimensional optical lattice. 
Section \ref{section:imaging} focuses on the imaging process and an analysis of the obtained resolution. 
In Sec. \ref{section:lifetime} we discuss the parameters that limit the lifetime in our ``deep-potential'' strategy. 
Finally, Sec. \ref{section:reconstruction} presents a short analysis of the lattice reconstruction process, where the original density distribution is determined from the obtained fluorescence images. This is followed by Sec. \ref{section:conclusion}, where we summarize and conclude our analysis.

\section{Experimental setup and lattice preparation}
\label{section:experiment}

\begin{figure}[t]
\includegraphics{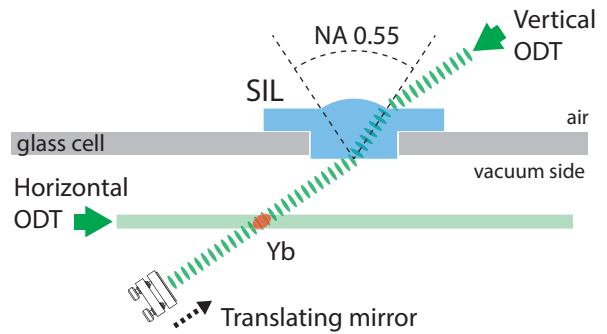}
\caption{\label{fig:intro} (Color online) Lateral view of the SIL attached to the 3-$\mathrm{mm}$-thick glass cell. Evaporative cooling is performed in a crossed ODT. }
\end{figure}

The experimental setup consists of a glass cell containing a solid immersion lens (SIL) connected to a metal vacuum chamber at a pressure of $2 \times 10^{-9}\,\textrm{Pa}$. The glass cell is made from 3-$\mathrm{mm}$-thick optical-quality glass, and the upper surface has a through hole with a diameter of 12$\,\mathrm{mm}$, where the SIL is attached (see Fig.~\ref{fig:intro}). The  SIL is a hemispherical lens made of fused silica glass, and it is divided into three parts. The upper part is a spherical cap with a radius of $8\,\mathrm{mm}$ and a height of $1.32\,\mathrm{mm}$, corresponding to a numerical aperture (NA) of 0.55. The middle part consists of a 3-$\mathrm{mm}$-thick disk with a diameter of $24\,\mathrm{mm}$, and the lower part is a 3.68-$\mathrm{mm}$-thick disk with a diameter of 11.5$\,\mathrm{mm}$. All three parts were optically contacted, and the lower flat and upper spherical surfaces were superpolished to reduce stray light. The SIL was glued to the glass cell using a vacuum leak sealant (SPI Vacseal) compatible with ultrahigh vacuum. The importance of the SIL here is that it increases the resolution of the microscope by a factor of 1.47 (the index of refraction of the fused-silica glass) \cite{mansfield:2615} and additionally allows us to create a stable optical system by fixing the position of the lattice with respect to the SIL flat surface \cite{PhysRevA.86.063615}.

Our experiment starts by loading ${}^{174}$Yb atoms into a magneto-optical trap and transferring the atoms into the glass cell using an optical tweezer, as explained in our previous experimental work \cite{PhysRevA.86.063615}. The horizontal optical dipole trap (ODT) beam transports 1.0$\times$10$^7$ atoms at a temperature of 25$\,\mathrm{\mu K}$ to a distance of 2.7$\,\mathrm{mm}$ below the surface of the SIL. A vertical ODT beam is then introduced, entering from the upper flat surface of the SIL and exiting at the center of the lower flat facet. The vertical ODT beam intersects the horizontal beam to form a crossed-beam ODT. Evaporative cooling is then performed by ramping down the intensity of the horizontal ODT beam in 4$\,\mathrm{s}$, resulting in 1.0$\times$10$^6$ atoms at a temperature of 2$\,\mathrm{\mu K}$. Vertical transport of the atoms close to the surface of the SIL is realized by adiabatically converting the vertical ODT into an optical ``conveyor belt'' and moving the retroreflection mirror (see Fig.~\ref{fig:intro}). Using this method, we can transfer $5.0 \times 10^5$ atoms at a temperature of $2\,\mathrm{\mu K}$ to a distance of $20 \,\mathrm{\mu m}$ below the surface of the SIL. It is important to note that the incident angle of the vertical ODT beam is set to the Brewster one. For this angle of incidence, the refracted angle is 34.2$^{\circ}$, which is slightly larger than the aperture of the spherical cap ($33.4^{\circ}$). This ensures that the beam will enter and exit from flat surfaces, thus avoiding any interference that may complicate the transport process.

\begin{figure}[t]
\includegraphics{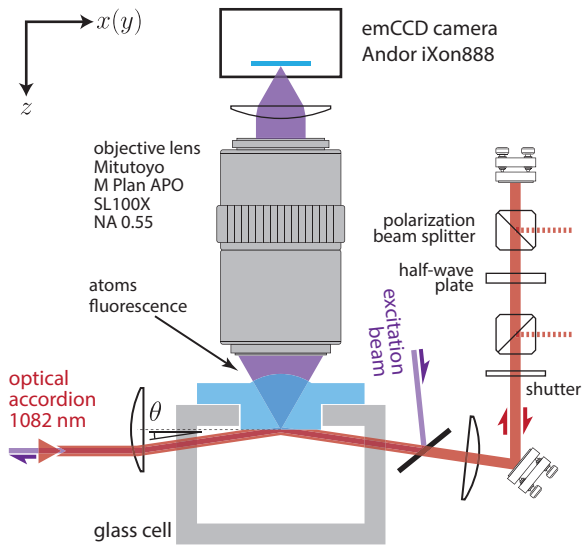}
\caption{\label{fig:experiment} (Color online) Experimental apparatus to trap and image atoms in a two-dimensional optical lattice. The microscope system is formed by an objective lens with a NA of 0.55 that, combined with a SIL, improves the NA of the system to 0.81. A retroreflected optical accordion is used to create the two-dimensional optical lattice. }
\end{figure}

After transporting the atoms, the optical accordion technique is used to create a two-dimensional BEC. This technique consists of reflecting a laser beam in a shallow angle onto the SIL, creating a standing wave with a periodicity dependent on the incident angle. The incident angle $\theta$ can be manipulated to compress the atomic cloud. Details of this technique are described in our previous report \cite{PhysRevA.86.063615}. For this experiment we replaced a single optical accordion with two orthogonal optical accordions aligned at the same incident angle, and shifted by a frequency of 100$\,\mathrm{MHz}$. Each of the beams has a wavelength of 1082$\,\mathrm{nm}$ and an elliptic cross section with a waist of 27 and 41$\,\mathrm{\mu m}$. 

To load the atoms into the accordion potential, the intensity of the optical accordion is gradually increased. while the vertical ODT is transformed back into a propagating wave by reducing the intensity of the retroreflected beam. As a result, atoms are transferred to a crossed trap between the two optical accordions and the vertical ODT. A condensate is then generated by evaporative cooling, first by reducing the power of the optical accordions and later by removing the vertical ODT completely, which resembles an evaporative cooling technique using a dimple trap \cite{jacob2011production}. The number of atoms in the condensate was 3.0$\times 10^4$ after 1.8$\,\mathrm{s}$ of evaporation. The accordion incident angle $\theta$ is then gradually changed from 0.7$^{\circ}$ to 6$^{\circ}$ in 250$\,\mathrm{ms}$ to compress the condensate into a pancake shape. After compression, 1$\times 10^4$ atoms remain in the trap.

Finally, atoms are loaded into the two-dimensional optical lattice by retro-reflecting each of the optical accordions. A system comprising two polarized beam splitters and a half-wave plate mounted on a hollow stepping motor is used to gradually increase the reflected beam in 70$\,\mathrm{ms}$ (see Fig.~\ref{fig:experiment}) and avoid heating of atoms.

Neglecting the Gaussian envelope of the laser beam and losses in the reflections, the final lattice has the following shape:
\begin{equation}
V_{\text{lat}} = -V_0 \left[ \cos^2 (k_x x) + \cos^2 (k_y y) \right] \sin^2 (k_z z),
\end{equation}
where $V_0$ is the lattice depth and $z=0$ is the surface of the SIL (see Fig.~\ref{fig:experiment} for the coordinate system). For an optical accordion with an incident angle of $\theta = 6^{\circ}$, the wavenumbers for orthogonal directions are  $k_x = k_y = 2 \pi \cos \theta / \lambda_{\text{lat}}$ and $k_z = 2 \pi \sin \theta / \lambda_{\text{lat}}$, corresponding to a lattice spacing of 544$\,\mathrm{nm}$ and 5.2$\,\mathrm{\mu m}$ in the horizontal ($x$, $y$) and vertical ($z$) directions, respectively. Initially, all the atoms in the pancake-shaped cloud are located in the first layer at $z = 2.6\,\mathrm{\mu m}$. The initial lattice depth is approximately $V_0/k_B=1.9$$\,\mathrm{\mu K}$ or $V_0/E_r=40$, with trap frequencies of $(\omega_x,\omega_y, \omega_z)/2\pi = (12,12,1.8)$ kHz.

\section{Imaging of atoms}
\label{section:imaging}

The imaging process starts by increasing the lattice depth to $V_0/k_B =150\,\mathrm{\mu K}$ ($3300 E_r$) in 10$\,\mathrm{ms}$. For this lattice depth, the potential depth on the ${}^1P_1$ state is $U_e/k_B=-68\,\mathrm{mK}$, corresponding to a relative light shift of $\Delta_0 /2\pi = (U_e - U_g) / h = -1.3$ GHz. In the case of an atom experiencing an average potential $(U_e+U_g)/2$, the corresponding trap frequencies are $(\omega_x,\omega_y, \omega_z)/2\pi = (1.6,1.6,0.2)\,\mathrm{MHz}$. The excitation light is aligned on the same optical axis as the optical accordion (see Fig.~\ref{fig:experiment}). This laser beam was derived from a frequency-doubled Ti:sapphire laser, and it has a power of 30$\,\mathrm{mW}$ and a beam waist of 100$\,\mathrm{\mu m}$, corresponding to a peak intensity of 95$\,\mathrm{W/cm^2}$ or $1600 I_s$, where $I_s$ is the saturation intensity of the ${}^1S_0$-${}^1P_1$ transition. The reason for using such a high-power density is that, as the excitation light is shone onto the lattice, atoms are rapidly heated. This causes the atoms to move from the center of the lattice site, which decreases the population in the excited state as the excitation beams become detuned. The excitation light with a high intensity alleviates this problem by broadening the spectrum of the ${}^1S_0$-${}^1P_1$ transition. 

\begin{figure}[t]
\includegraphics{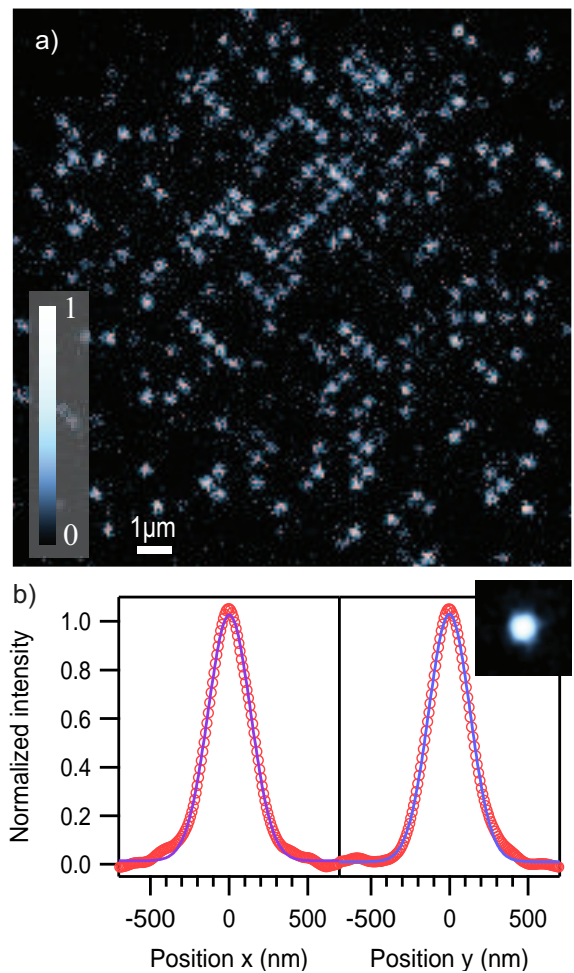}
\caption{\label{fig:result} (Color online). (a) Raw fluorescence image of atoms trapped in a two-dimensional optical lattice. The false-colored scale is in arbitrary units. (b) Profile of 30 averaged isolated sites (inset image) and the corresponding intensity profile along the center for each direction. Both curves were fitted using a Gaussian profile, resulting in a resolution (FWHM) of 318(2)$\,\mathrm{nm}$ and 310(2)$\,\mathrm{nm}$ for the $x$ and $y$ directions, respectively.}
\end{figure}

After 100$\,\mathrm{\mu s}$ of exposure time, a single-shot image of the Yb atoms trapped in a two-dimensional optical lattice was successfully observed [Fig.~\ref{fig:result}(a)]. To measure the point spread function (PSF) and the resolution of the system, 30 images of isolated sites were averaged, and the intensity profile along each axis was fitted using a Gaussian function [Fig.~\ref{fig:result}(b)]. The resultant FWHM resolution was 318(2) and 310(2)$\,\mathrm{nm}$ in the $x$ and $y$ directions, respectively. For a microscope system with a NA of 0.81 and an imaging wavelength of $\lambda= 399\,\mathrm{nm}$, the theoretical resolution based on the Rayleigh criterion is $0.51\lambda / \text{NA} = 250\,\mathrm{nm}$. The difference between the experimentally obtained resolution and the theoretical one might be related to the fact that the atoms are not completely at rest during the imaging process.

Note that the exposure time (100$\,\mathrm{\mu s}$) using the technique described in this paper is four orders of magnitude smaller than that of previous reports using alkali atoms \cite{nature08482, nature09378}, where exposure times on the order of 1$\,$s are used. The total number of collected photons is also one order of magnitude smaller than the one reported in \cite{nature08482}. Even under such circumstances, we can resolve single sites in a single shot with excellent signal-to-noise ratios. One of the advantages of having a short imaging time is that the system is robust against mechanical instabilities in the optical system. 

\begin{figure*}[t]
\includegraphics[width=185mm]{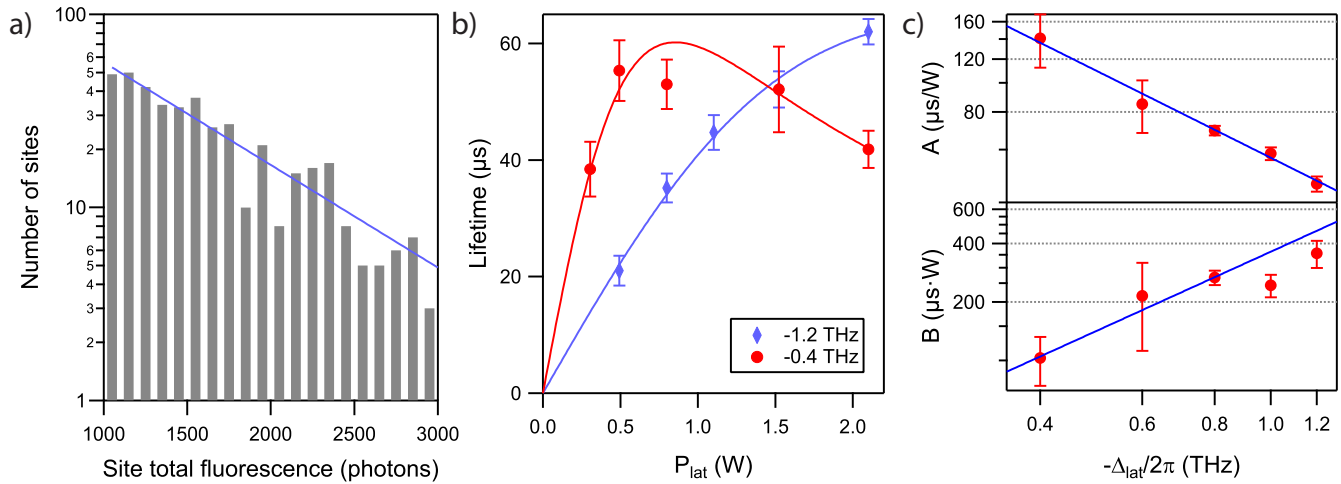}
\caption{\label{fig:lifetime} (Color online) Lifetime of atoms during imaging process. (a) Lifetime for $\Delta_{lat}/2\pi=-0.6\,\mathrm{THz}$ and $P_{lat}=1.1\,\mathrm{W}$ is calculated by fitting the site-intensity histogram with an exponential decay and dividing the result by the photon emission rate. (b) Lifetime dependency on lattice intensity $P_{lat}$ for $\Delta_{lat}/2\pi=-0.4\,\mathrm{THz}$ and $\Delta_{lat}/2\pi=-1.2\,\mathrm{THz}$. (c) Dependency of parameters $A$ and $B$ on lattice detuning $\Delta_{lat}$. Each parameter result was fitted with a power function. }
\end{figure*}

\section{Lifetime}
\label{section:lifetime}

During the imaging process, atoms absorb the excitation light and are heated up due to random light scattering. After a number of scatterings, the atom escapes from the site and is lost. The lifetime of the atoms was determined using the histogram of the total intensity in each site. Assuming that the number of photons emitted by each atom is proportional to the lifetime, and the number of atoms decreases exponentially, the shape of the histogram is also expected to decay exponentially for imaging times much longer than the lifetime. 

Photoassociation (PA) effects in multiply occupied sites result in the formation of molecules that lead to the loss of atoms in times of the order of a few to tens of microseconds \cite{PhysRevLett.91.080402}. As the short imaging times in our system are comparable with these times, PA losses cannot be neglected. Multiply occupied sites with their associated PA effects could modify the shape of the histogram and affect the lifetime estimation. Here, we opted to use very sparsely occupied (filling $< 0.05$) lattices to reduce these effects.

To understand what parameters limit the lifetime in our system, we studied the change in the lifetime for different lattice powers $P_{lat}$ and lattice detunings $\Delta_{lat}$. Figure \ref{fig:lifetime}(a) shows a typical histogram result for $P_{lat}=1.1\,\mathrm{W}$ and $\Delta_{lat}/2\pi = -0.6\,\mathrm{THz}$. The lifetime is obtained by dividing the result of the exponential fitting by the photon emission rate. This rate was experimentally estimated as 13.0(3)$\,\mathrm{photons/\mu s}$, knowing that an average atom emits 520(10) photons for an exposure time of 40$\,\mathrm{\mu s}$.

We consider two possible causes that limit the lifetime in the system. The first cause is the heating due to the excitation beam scattering. As the kinetic energy of the atom is approximately proportional to the number of emitted photons, the lifetime $\tau_1$ due to heating is proportional to the potential depth of the lattice. In other words, the lifetime $\tau_1$ is proportional to the lattice power $P_{lat}$. The second possible cause is the absorption of the optical lattice light, which excites the atoms to the (6$s$7$s$)${}^1S_0$ state. As the potential in this state is strongly repulsive and the decay rate from it is comparable to the trap frequency, we assume that the atom will be lost after an excitation. Consequently, the lifetime $\tau_2$ due to the absorption of the lattice light is proportional to $1/P_{lat}$. The total lifetime $\tau$ can then be written as
\begin{equation}
\label{eq:lifetime1}
\tau = \left( \frac{1}{\tau_1} + \frac{1}{\tau_2} \right)^{-1}.
\end{equation}
We estimate the dependence of the lifetime on the lattice power for a constant lattice detuning. Figure \ref{fig:lifetime}(b) shows the measured lifetimes as a function of the lattice power $P_{lat}$ for two different lattice detunings $\Delta_{lat}/2\pi = -0.4\,\mathrm{THz}$ and $\Delta_{lat}/2\pi = -1.2\,\mathrm{THz}$. Solid lines are the fitted curves using Eq.~(\ref{eq:lifetime1}) with $\tau_1 = A P_{lat}$ and $\tau_2 = B / P_{lat}$, where $A$ and $B$ are two fitting parameters dependent on the lattice detuning. 

The same analysis is repeated for five different detunings, from which we obtain the $A$ and $B$ parameters for each. Each of the parameters is also fitted using the power equations $A = a {\Delta_{lat}}^\alpha$ and $B = b {\Delta_{lat}}^\beta$, resulting in $\alpha = -0.96(13)$ and $\beta = 1.4(4)$, as shown in Fig.~\ref{fig:lifetime}(c). If only the strong attractive potential in the ${}^1P_1$ state determines $A$, we expect that $A \propto {\Delta_{lat}}^{-1}$ for a two-level system under the rotating-wave approximation. In addition, if only the scattering process due to the lattice light determines $B$, we also expect that $B \propto {\Delta_{lat}}^{2}$. Both of the values of $\alpha$ and $\beta$ are consistent with these rough estimates within two standard deviations.

For the maximum lattice power  $P_{lat} = 2.1\,\mathrm{W}$ in our system, we obtained a lifetime of 62(2)$\,\mu$s at $\Delta_{lat}/2\pi = -1.2$$\,\mathrm{THz}$. Using the obtained fitting results, it is also possible to estimate the maximum lifetime for a given power $P_{lat}$. Upgrading the system to achieve a maximum output of $P_{lat} = 10\,\mathrm{W}$ would increase the lifetime to 92$\,\mu$s at $\Delta_{lat}/2\pi = -3.8$$\,\mathrm{THz}$ ($\lambda_{lat} = 1092\,\mathrm{nm}$). Fortunately, high-power laser sources are readily available for this wavelength.

\section{Lattice reconstruction}
\label{section:reconstruction}

In this section, we analyze the reconstruction process that determines the original density distribution of atoms in the optical lattice from each of the obtained fluorescence images. To reduce the effects of PA and hopping we will limit the discussion to sparsely filled lattices (filling $\sim$0.1). The experiments are realized with an exposure time of $40\,\mathrm{\mu s}$, a lattice power of $P_{lat} = 2.1\,\mathrm{W}$, and a lattice detuning of $\Delta_{lat}/2\pi = -1.2$$\,\mathrm{THz}$. 

\begin{figure}[t]
\includegraphics[width=77mm]{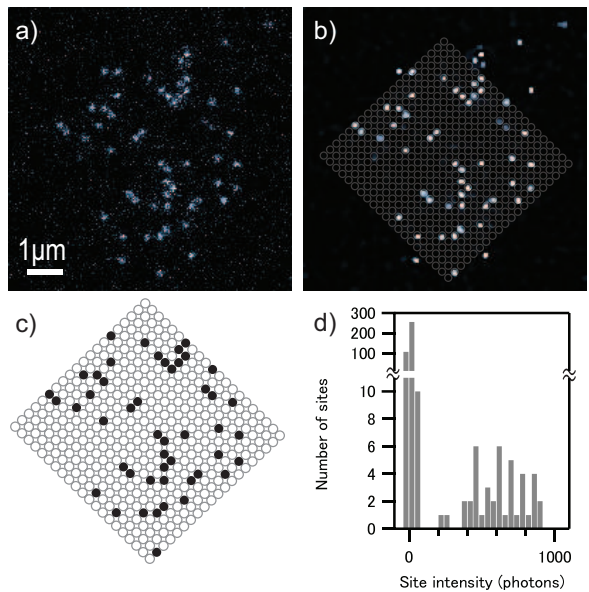}
\caption{\label{fig:histogram} (Color online) Reconstruction of the original density distribution of atoms in a two-dimensional optical lattice. (a) False-colored original image. (b) Deconvoluted image of (a) using the PSF obtained in Fig.~\ref{fig:result}(b). (c) Site occupancy when setting a threshold at 80 photons. Black and white circles represent occupied sites and empty sites, respectively. (d) Histogram of the photon counts in each site, for the image in (a)-(c). }
\end{figure}

Our algorithm first deconvolutes the fluorescence image using the PSF obtained in Fig.~\ref{fig:result}(b), and determines the best geometry (angle, periodicity, and phase) for the optical lattice that maximizes the total local intensity in each site. The result of this process is shown in Figs.~\ref{fig:histogram}(a) and \ref{fig:histogram}(b). The site total intensity is then obtained by binning the pixels in each site. Figure \ref{fig:histogram}d shows a typical histogram that shows a clear double-peaked shape representing empty sites and singly occupied sites. 

To determine whether a site is occupied or not, a threshold is set at 80 photons, resulting in the atom density distribution shown in Fig.~\ref{fig:histogram}(c), where the black circles represent occupied sites. 

The accuracy in determining empty sites can be estimated by knowing the histogram profile in the absence of atoms. From this profile we can estimate that the algorithm correctly determines empty sites in $>\!99\,\mathrm{\%}$ of the cases. The determination of occupied sites is highly influenced by atom losses and hopping effects. For a lifetime of 62$\,\mu$s (800 photons), $10\,\mathrm{\%}$ of the atoms are lost before emitting the 80 photons corresponding to the threshold. Such sites would be incorrectly estimated as empty sites, greatly decreasing the accuracy of the estimation. The usual method to evaluate hopping effects is to take multiple images from the same lattice and verify whether each atom position has changed or not. In our experimental setup the lifetime and exposure time are very short, and it is technically difficult to utilize this method. However, we expect that hopping effects are greatly reduced, as the atoms that escape from the site and hop to neighboring sites have a high kinetic energy of approximately 30$\,\mathrm{mK}$ and cannot be cooled by the excitation beam. Such atoms are expected to hop indefinitely until they are lost, emitting a reduced number of photons in the neighboring sites. As an alternative way to estimate hopping effects we measured the intensity correlation function $g^{(2)}(R)=\langle I(r)I(r+R) \rangle / \langle I(r)^2 \rangle$ of occupied sites in 200 sparse lattice images. This resulted in correlations of $(1.03(18),1.03(24),1.03(19))$ for interatomic distances $R=(1,\sqrt{2},2)\times 544$$\,\mathrm{nm}$. These correlation values indicate that hopping events do not produce enough photons to be counted as an occupied site.

\section{Conclusions}
\label{section:conclusion}

In conclusion, we have demonstrated a high-resolution microscope system capable of resolving individual Yb atoms trapped in a two-dimensional optical lattice with single-site resolution. Our system utilizes a deep potential in the excited state to trap the atoms that are heated during the imaging process, requiring no additional cooling method. We studied the lifetime limitations of this system and showed an algorithm that reconstructs the original atom density distribution. The experimental setup used to load and image atoms in a two-dimensional optical lattice does not depend on the energy level structure and is completely all optical and thus can be adapted to other Yb isotopes or atomic species.

\begin{acknowledgments}
We would like to thank K. Aikawa and M. Ueda for their stimulating and fruitful discussions. This work is supported by a Grant-in-Aid for Scientific Research on Innovative Areas ``Fluctuation \& Structure'' from MEXT, 
the Cabinet Office, government of Japan, through its ``Funding Program for Next Generation World-Leading Researchers'';
the Matsuo Foundation; the Murata Science Foundation; and the Research Foundation for Opto-Science and Technology.
One of the authors (M.M.) is supported in part by the Japan Society for the Promotion of Science.
\end{acknowledgments}

\bibliography{second}

\end{document}